\documentclass[prd,aps]{revtex4}

\begin{document}

\title{The ADM papers and part of their modern legacy: loop quantum gravity}

\author{Jorge Pullin}
\affiliation {
Department of Physics and Astronomy, Louisiana State University,
Baton Rouge, LA 70803-4001}

\begin{abstract}
We present a summary for non-specialists of loop quantum gravity as
part of the modern legacy of the series of papers by Arnowitt, Deser
and Misner circa 1960.
\end{abstract}

\maketitle
\section{Introduction: the ADM papers}

The revolutionary character of Einstein's general theory of relativity
can be gauged by how long it took to appreciate fully its
implications. Although the equations of the theory and some basic
consequences were laid down in 1916, many of the concepts and ideas
that we consider important today would have to wait several decades
for their development. For instance, although the Schwarzschild
solution was written also in 1916, the idea of black hole was not
properly understood until the 1960s. It is remarkable that many of the
brightest minds of the 20th century physics missed the concept or were
quite confused by it. Gravitational waves, already discussed by
Einstein himself in 1916, had a tortuous development (with Einstein
himself coming to doubt their existence) until the binary pulsar put
to rest the confusion about their existence \cite{kennefick}. 

Another of the concepts that remained in a confused state for many
years was the dynamics of Einstein's theory. This is not
surprising. Einstein's equations are a complex set of non-linear
partial differential equations. Unless handled properly, their
character is unclear since they mix wave like equations with elliptic
equations. This created some confusion even as recently as the 1990's
in the context of numerical relativity. Mainly, the Einstein equations
are equations for a space-time and we are accustomed to describe
physics as an evolution in time. However, general relativity is a
background independent theory, implying that there is no preferred
notion of time and therefore not a preferred notion of evolution. This
was the source of confusion over the years. Early work on
understanding the Einstein equations as an initial value problem was
done by Darmois and Lichnerowicz (see the nice recent account of
Choquet-Bruhat \cite{choquet}). Experience with other theories
suggested that further insights could be obtained by casting the
theory in canonical form. However, the elaborate symmetries of general
relativity meant that progress in the canonical formulation was not
achieved till the late 1950's by Dirac \cite{dirac} and independent
developments were done by Bergmann's group in Syracuse \cite{bergmann}.

Then, in an illuminating series of papers around the year 1960,
Arnowitt, Deser and Misner (ADM) \cite{admseries} provided remarkable
insights concerning the canonical formulation and the
identification of the true degrees of freedom, using techniques
similar to those Schwinger had used in other field theories.  This
body of work led to ADM receiving the Dannie Heinemann Prize of the
American Physical Society in 1994. Richard Arnowitt unfortunately
passed away last year. The Einstein medal has recently been awarded to
Stanley Deser and Charles Misner also citing the same body of work.

A good summary of the results is still the paper by Arnowitt, Deser
and Misner \cite{admwitten} in the volume edited by Louis Witten that
presented an interesting snapshot of current research in relativity in
the early 1960's. This summarizes a series of papers by the same
authors in journals. It was reprinted as a ``Golden Oldie'' by the
journal General Relativity and Gravitation \cite{goldenoldie}. A
recent paper on the legacy of ADM was written by Stanley Deser \cite{deseradm}.

The paper starts by recalling and casts in simple notation the
Hamiltonian framework for parameterized systems, where the Hamiltonian
is a constraint. Lanczos \cite{lanczos} already discusses this idea,
which goes back to Jacobi, but the presentation in this paper is
cleaner, almost as one would present it today. The idea is that in
parameterized mechanical systems the action takes the form
\begin{equation}
  S= \int dt \left( \sum_{i=1}^M p_q \dot{q}_i -N H\left(q_i,p_i\right)\right),
\end{equation}
where $N$ is a Lagrange multiplier and varying with respect to it yields 
$H(q_i,p_i)=0$, a constraint. The next section of the paper goes on to
show that the action of general relativity can be written in a way
that it is manifestly in ``already parameterized'' form. 

Then the paper recalls the Palatini formulation of general relativity,
where the connection is treated as an independent variable and writing
the Ricci tensor as a function of the connection only,
\begin{equation}
  S= \int d^4x \sqrt{-q} g^{\mu\nu} R_{\mu\nu}(\Gamma),
\end{equation}
so variation with respect to $g^{\mu\nu}$ yields the Einstein
equations and variation with respect to the connection gives the
Christoffel formula. 
It then proceeds to consider a $3+1$ decomposition of the metric 
\begin{eqnarray}
g_{ij}&=&{}^4g_{ij},\qquad N=\left(-{}^4g^{00}\right)^{-1/2},\qquad N_i={}^4g_{0i}\\
\pi^{ij}&=&\sqrt{-{}^4g}\left({}^4\Gamma^{0}_{kl}-g_{kl}{}^4\Gamma^0_{mn}g^{mn}\right)g^{ik}g^{jl},
\end{eqnarray}
with $i,j,k...$ going from one to three, and rewrites the Lagrangian
as 
\begin{equation}
  {\cal L}= g_{ij}\partial_t \pi^{ij} -N H -N_i H^i+{\rm div}
\end{equation}
with ${\rm div}$ a divergence term and 
\begin{eqnarray}
  H&=&\sqrt{g} \left[{}^3R+g^{-1} \left(\frac{1}{2} \pi^2
      -\pi^{ij}\pi_{ij}\right)\right],\\
H^i&=& -2 \pi^{ij}_{\vert j},
\end{eqnarray}
where we recognize that it is already in a parameterized form as
discussed previously and identify the Hamiltonian $H$ and
diffeomorphism $H^j$ constraints. It goes on to notice that one can
work out from the Lagrangian the equations of motion for $g_{ij}$ and
$\pi^{ij}$ and together with the constraints, they are equivalent to
the Einstein equations. The paper discusses the meanings of the
various variables, how they are intrinsic to a surface $t={\rm
  constant}$ and the roles of the multipliers $N$, $N^i$ and the
interpretation of the momentum $\pi^{ij}$ and its relation to the
extrinsic curvature of the surface.

The paper also uses the constraint structure and the Lagrange
multipliers to elucidate the initial value problem of general
relativity. It makes clear that if one specifies $g_{ij}$, $\pi^{ij}$,
$N$ and $N^i$ at a given time $t$, the equations determine the values
of $g_{ij}$ and $\pi^{ij}$ at a later time. However, $g_{ij}$ and
$\pi^{ij}$ cannot be chosen arbitrarily at an initial time $t$, they
are subject to constraints. It also notes that the constraints are
preserved in time due to the Bianchi identities and therefore are
satisfied at all future times if they are satisfied initially. 

Moreover it is pointed out that the twelve dynamical variables
$g_{ij},\pi^{ij}$ provide a complete Cauchy data set but not a minimal
one. That will be further elucidated in the canonical section of the paper.

Then it casts the theory in canonical form, and starting from an
analogy with linearized theory proceeds to fix the gauge of the theory
to have a true Hamiltonian dictate the evolution. Having the true
Hamiltonian it can proceed to define what we now call the ADM energy
and ADM momentum of the gravitational field,
\begin{eqnarray}
  E&=& \oint dS_i \left(g_{ij,j}-g_{jj,i}\right),\\
  P^i &=& -2 \oint \pi^{ij} dS_j.
\end{eqnarray}
These, together with the initial value formulation and the gauge
fixings are widely used in numerical relativity, perhaps the other
main legacy of the ADM papers. We will not cover the topic here but
refer to the article by Sperhake \cite{sperhake} in this volume. 

The paper then goes on to talk about gravitational radiation, in
particular the definition of the wave-zone, a delicate concept in a
non-linear theory. This should be put in the context that at the time
there was controversy in certain circles on the existence of
gravitational radiation (see Kennefick for a complete account of the
controversy). It also considers coupling the theory to
electromagnetism. 

The paper concludes with some discussions on self-energy, in
particular including an illuminating example about how non-perturbative effects
could influence the self-energy. This example is cited in the
introduction to one of the first books in loop quantum gravity by
Ashtekar and Tate as motivation \cite{ashtekartate}. 

Attempts to use the canonical formulation described in this paper to
quantize general relativity were done in the 1960's. The idea was to
consider wavefunctions $\Psi[g]$ that depended on the spatial metric
(e.g. \cite{dewitt}). The canonical momenta operated like functional
derivatives and the constraints promoted to operatorial
equations. These ran into several difficulties. Among them there was
no success in constructing an inner product that would constitute a
Hilbert space and that was compatible with the symmetries of the
theory. There was little mathematical control on the space of
wavefunctions of the metric. And the complicated non-polynomial nature
of the Hamiltonian constraint made it difficult to promote it to a
well defined quantum operator. Even in simplified contexts like
quantum cosmology, where one freezes all degrees of freedom but a
finite number of them, the resulting quantum theories failed to
achieve noteworthy results, in particular, the elimination of the
singularity that appears in the classical theory. This led to a
slowing of activity in canonical quantum gravity in the mid 1970's.

\section{Loop quantum gravity: the beginnings}

In 1986 Ashtekar \cite{ashtekar} presented a reformulation of the
canonical treatment of general relativity. Its original presentation
was in terms of spinors, and part of the original calculations and
motivations also came from spinors, but it turns out they are not essential
to present the theory.  The reformulation consisted of replacing the
spatial metric with a (densitized) triad $\tilde{E}^a_i$,
where $a$ corresponds to a spatial index and $i$ is a triad index that
is raised and lowered with a flat ($SO(3)$ invariant) metric. The
canonically conjugated momentum is related to the extrinsic curvature
and the spin connection associated with the triad $A_a^i$. Such a
variable had been invented by Sen in a different context
\cite{sen}. The novelty is that it transforms as an (complex) $SO(3)$
connection. These are the same variables one would use to describe an
$SO(3)$ Yang--Mills theory. Due to the additional number of variables
there exist an additional set of constraints that are identical to the
Gauss law that one has in Yang--Mills theories. Therefore one could
view general relativity as a theory with the same phase space as
Yang--Mills theories with additional constraints. The constraints took
a very simple polynomial form,
\begin{eqnarray}
  D_a \tilde{E}^a_i&=&0\\
  \tilde{E}^a_i F_{ab}^i&=&0\\
\epsilon_{ijk} \tilde{E}^a_i\tilde{E}^b_j F_{ab}^k = 0,
\end{eqnarray}
where $D_a$ is the covariant derivative and $F_{ab}^k$ the curvature
of the connection $A_a^i$. 
The new variables immediately led to a shift in point of view: the
natural polarization one uses to treat Yang--Mills theories is to
choose wavefunctions of the connection $\Psi[A]$, not of the electric
field. That choice would be equivalent to considering
wavefunctions of the extrinsic curvature rather than the metric. The
wavefunctions have to be gauge invariant functions under $SO(3)$
transformations plus satisfy four additional constraints. All
this led to the hope that the many successful techniques that were
used to treat Yang--Mills theories could be imported into gravity. In
the end the latter hope proved naive: even if one could cast them as
sharing an (unconstrained) phase space, gravity and Yang--Mills
theories are vastly different theories. Some of the reasons that
Yang--Mills theories can be treated, such as their asymptotic freedom in
the case of QCD, and the possibility of putting them on a lattice are
just not available in the gravitational case. One technique, however,
proved fruitful.

The technique in question is the use of loop variables. The key idea
is contained in the familiar Stokes' theorem,
\begin{equation}
  \int_{\partial S} \vec{A}\cdot d\vec{\ell} =\int_S \vec{B}\cdot \vec{dS},
\end{equation}
where $\partial S$ is a loop given by the boundary of the surface
$S$. Given a vector potential, the left hand side of this equation is
a function of a loop. If one knows the value
of that function for all loops, for a vector potential, one knows
the magnetic field. A simple way of seeing this is to consider
infinitesimal loops where the right hand side is just the value of the
field times the area of the infinitesimal loop. A similar result holds
for the vector potentials (connection) of Yang--Mills theories. So
this opens the possibility to replace the wavefunctions of the
connection $\Psi[A]$ with functions of loops $\Psi[\gamma]$ with
$\gamma$ a closed curve. In the context of Yang--Mills theories (and
even gravity \cite{fustero}) this was first studied by Gambini and
Trias \cite{gambinitrias}. In the context of gravity written in terms
of Ashtekar's new variables this was first studied by Rovelli and
Smolin \cite{rovellismolin}. The resulting quantum representation is
known as the loop representation. One can connect the connection
representation with the loop representation via a ``loop transform'',
\begin{equation}
  \Psi[A]=\int d\gamma W_\gamma[A] \Psi[\gamma]
\end{equation}
where the integral is a formal sum over loops (which has been made
precise \cite{thiemanntransform}) and $W_\gamma[A]$ is the trace of the holonomy of the
connection $A_a^i$ along the loop $\gamma$, the non-Abelian
generalization of the circulation of the vector that appears in
Stoke's theorem.

Again, the loop representation does not create miracles, but it shifts
the point of view on several problems. First of all, the loop
variables are gauge invariant, so the Gauss law is automatically
satisfied. Secondly, there is significant experience in writing
functions of loops that are invariant under diffeomorphisms, it is the
branch of mathematics known as knot theory. So one was left with the
task of understanding the Hamiltonian constraint.

Several attempts were made to write the Hamiltonian constraint in
terms of loop variables in the early 1990's \cite{hamiltonian90s}. It
helped that the structure of the constraints in terms of the new
variables is polynomial. However, the expression still had to be
regularized, which is not easy to do in a diffeomorphism invariant
way. Moreover, the loop variables are not free, they are constrained
by a set of non-linear identities known as Mandelstam identities
\cite{mandelstam}. Also, the complex nature of the Ashtekar variables
(one was using complex coordinates to describe a real theory) required
imposing extra ``reality conditions'' on the framework. These problems
hampered progress until new elements were put into place that address
them.

\section{The 1990's: quantum geometry and a well defined theory of
  quantum gravity}

The situation started to improve when in 1995 Rovelli and Smolin
\cite{rovellismolinspinnets} noted that one could use spin networks to
solve the Mandelstam identities and provide a basis of (almost)
independent loop states. Spin networks are graphs with intersections
and ``colors'' on their lines that had presciently been studied by
Roger Penrose \cite{penrosespinnets} in the late 1960's as potentially
connected with quantum gravity. In terms of these states it was
possible to write clean expressions for the operators representing the
area of a surface and the volume of a region of space. A picture of
quantum geometry started to emerge: surfaces were endowed with area
when punctured by the lines of the spin networks and regions of space
were endowed with volume when they contained vertices of the spin
networks \cite{ashtekarlewandowskiareavolume}.

Also in the same year, Ashtekar and Lewandowski
\cite{ashtekarlewandowski} showed how one could use spin networks to
introduce a diffeomorphism invariant inner product in the space of
connections modulo gauge transformations. The resulting inner product
is remarkable as there were no known inner products in such a space
known in closed form (Yang--Mills theory provides an inner product
perturbatively or on the lattice). The resulting inner product had a
particularly simple form in terms of spin network states. Two states
are orthogonal if their graphs or ``colors'' are different. Otherwise
their inner product is unity.

This inner product leads to unusual consequences. In particular it
emerges that there is no well defined operator corresponding to the
connection in the resulting Hilbert space. What has a well defined action
is the holonomy. This may appear strange since the latter is the path
ordered exponential of the former, but there is no well defined notion
of ``logarithm'' to invert the definition. This in particular implies
a departure from the type of functional spaces that are normally
considered in quantum field theory. Some people have questioned the
suitability of these spaces, in particular their non-separable nature
\cite{nicolai} (see response here \cite{thiemannresponse}). But it is
clear that if something new was going to be obtained in quantum
gravity one would have to relinquish structures that failed to do the
job in the past. We will see that even in the simple context of
cosmology this relinquishing has important consequences in the next
section.

Then in 1996 Thomas Thiemann \cite{qsd} surprised the community when
he announced that using the quantum geometry techniques he could write
a well defined quantum expression for the Hamiltonian constraint. The
expression was free of infinities and free of anomalies. It also could
be extended to matter couplings. Also, it did not use the original
Ashtekar variables (which are complex) but a real modification that
Barbero \cite{barbero} had put forward (more precisely a one-parameter
family of modifications, the parameter is known as Immirzi parameter
\cite{immirzi}). So it did not have to deal with reality
conditions. This was the first non-trivial, finite, well defined
theory of quantum gravity. The remaining question, unsolved until
today, is if it captures enough of the right physics of general
relativity in the semi-classical limit.

In a separate development, Rovelli and Krasnov \cite{krasnovrovelli}
in 1996 noted that one could make sense of the entropy of black holes
in loop quantum gravity by counting the number of possible spin
networks configurations that would endow the horizon with a given area
by piercing it. The result was further formalized by Ashtekar, Baez,
Corichi and Krasnov \cite{asbacokr}. They noted that if one formulated
the action classically on a manifold with a boundary given by the
horizon, a Chern--Simons theory arises on the horizon and certain
compatibility conditions have to be met at the ``punctures'' the spin
network generates on the horizon. The resulting entropy is
proportional to the area, but the proportionality factor depends on
the Immirzi parameter, whose value at the moment is not known. The
status of the Immirzi parameter has parallels with the theta ambiguity
in Yang--Mills theory. At the moment there are no other experimental
predictions that could determine the parameter. At least for all
calculations of black hole entropies including rotating, charged,
deformed and other black holes the dependence on the Immirzi parameter
is always the same, so a single value is compatible with them. The
subject of black hole entropy in loop quantum gravity has been further
developed by researchers in Spain \cite{spain} who used advanced
combinatorial techniques to do the counting of configurations and
discovered interesting structures in how eigenvalues for the area
cluster. In particular they can compute higher corrections in terms of
the area. There has been some dispute about the fact that the
logarithmic corrections do not seem to agree with those worked out by
more conventional Euclidean path integral techniques
\cite{ashokesen}. Many however share the view that it is unclear if
the two calculations are computing the same thing since they are so
different in nature. For instance, it should be remembered that in the
Euclidean context there really is no horizon.

Another separate development was the observation that the discrete
nature of quantum geometry could leave an imprint on the light that
arrives from distant gamma ray burts \cite{gammaray}. The discrete
nature of space-time leads to dispersion of the light. The effect is
minute, of the order of the Planck length divided by the wavelength,
but for gamma ray bursts the light travels a very long distance so the
effect can build up. It turns out the detailed (yet heuristic)
calculation shows that the order of the effect in inverse powers of
the wavelength depends on the parity chosen of the quantum state
representing the universe. Only if there is a parity violation one
would have a linear effect. And if the linear effect is absent then
the effect becomes too small to be measured with current
observations. It turns out that radio sources already put a stringent
limit on the linear effects \cite{gleiserkozameh}, suggesting that the
state describing space time does not violate parity. Many papers have
been written on these issues, a good summary is in \cite{summarygrb}.

\section{The 21st century: symmetry reduced models}

An avenue that was chosen by several researchers to try to probe the
physical content of Thiemann's theory is to consider symmetry reduced
models. In them one freezes most of the degrees of freedom of the
theory keeping only a small number of them dynamical, perhaps a finite
number of them. The resulting models usually are too simple to apply
``loop quantum gravity'' techniques to them, but one can use
techniques inspired by loop quantum gravity. 

A first set of such models comes from cosmology, where the first clean
treatment was carried out by Bojowald \cite{bojowald}. One may 
consider a Friedmann--Robertson--Walker model where the only degree of
freedoms in the metric is the scale factor $a(t)$. If one works out
the Ashtekar variables for that model, a gauge can be chosen in which
the connection takes the form $A_a^i= c(t) \delta_a^i$ and the triad
is related to the canonically conjugate variable to $c(t)$, usually
called $p(t)$. Notice that the resulting model is a mechanical system,
it has one degree of freedom, it is not a field theory. One then
proceeds to quantize, however, mirroring the full theory, one
constructs a Hilbert space that has some discreteness in it. Namely,
instead of the usual inner product $\langle p'\vert
p\rangle=\delta(p'-p)$ one writes $\langle p'\vert p\rangle=
\delta_{p',p}$ with a Kronecker delta. This is tantamount to saying
that the variable $c$ is not well defined but its exponentiation is,
$\sin(\rho c)/\rho$. Here $\rho$ is a free parameter that can be
viewed as the remnant information of the loop, i.e. its length, in the
context of a homogeneous space. In the limit $\rho\to 0$ one recovers
the usual theory, but such limit is not available in the Hilbert space
chosen inspired in the Ashtekar--Lewandowski measure. This implies
that the quantum evolution equations one writes will depart from the
classical behavior when $\sin(\rho c)\sim 1$. This occurs near where the big
bang is in the classical theory. The picture that emerges is that of a
universe that when it is large, is well approximated by general
relativity but if one runs it backwards, near where the big bang used
to be it ``bounces'' back into expansion into the past without a
singularity developing.  Some refinements on the picture were made
noticing that the role of $\rho$ was related to the minimum value of
the quantum of area around which to run a loop and therefore ought to
depend on the dynamical variables. This led to an ``improved
dynamics'' by Ashtekar, Pawlowski and Singh \cite{aps} that has by now been
applied in a variety of models and it appears that the feature of the
bounce appears in all the models considered. 

An important development has been the study of perturbations of the
loop quantum cosmology models in search of potential imprints of the
dynamics in the cosmic microwave background \cite{agulloetal}. At
first this may seem unlikely, the background gets constituted after
inflation and during it one expects quantum gravity corrections to be
of the order of $10^{-12}$. Although this is true, the dynamics of the
bounce influences the initial state that is evolved through inflation
and eventually gives rise to the cosmic microwave background.  The
result is that imprints are left in the very long wavelength modes
that could potentially be observable. Unfortunately the observational
data is very inaccurate for long wavelengths. Moreover the result
depends on the value of the inflaton at the bounce, which we do not
know. So it is not an unambiguous prediction of the theory that can be
used to confirm it or rule it out if it is not observed, one simply
can adjust the value of the inflaton at the bounce to match the
observed data. The ``tilt'' of the tensor spectral index, and its
consistency relation however, differ from those
predicted by standard inflation \cite{agulloetal}. This has not been
measured yet, but it is plausible that it could be measured in the
forthcoming years, leading to a prediction that can be used to confirm
or rule out the theory. The fact that contact with experiment is so
close is really an exciting possibility.

Another set of symmetry reduced models where the theory has been
applied has been spherically symmetric models. The initial treatment
was due to Bojowald and Swiderski \cite{boswi}. There one faces more
challenges than in the cosmological setting. The variables have
spatial dependence and therefore one in principle is dealing with
field theories (even though at the end of the day the number of
degrees of freedom may end up being finite in some cases, for instance
in vacuum). One has a diffeomorphism and Hamiltonian constraint and
one has to worry about their constraint algebra. Remarkably, it was
found that for several of these models one can perform linear
combinations of the constraints to yield a Hamiltonian constraint that
is Abelian. One can then complete the Dirac quantization of them
\cite{prl}. Remarkably, the constraints can be solved in closed
form. One can write parameterized Dirac observables acting on the
space of physical states that represent the components of the
metric. The parameters play the role of choices of gauge. The
resulting operator for the metric, in order for it to be self adjoint,
does not have support on the point where the singularity used to
be. So the black hole singularity is also resolved, and one can move
through the region where it was into another region of space-time
isometric to the exterior. Quantum fields have been studied on this
quantum space-time background and Hawking radiation has been recovered
with small corrections, as expected, and coinciding with a heuristic
calculation that had simply assumed a cutoff a few years ago
\cite{hawking}. The Casimir effect between two shells living on the
quantum space-time was computed \cite{casimir}, obtaining the correct
result without having to resort to regularization nor renormalization,
therefore opening new perspectives on the problem of back-reaction,
which now appears treatable.

\section{Other developments and conclusions}

Given the brevity of this article, we cannot do justice to many other
developments that have spun off from loop quantum gravity. The main
one perhaps is the development of a covariant form of the theory using
the path integral. The idea is to use the quantum geometry tools to
address some of the usual problems in the definition of the path
integral. The resulting theories are called spin foam models. There is
now a book by Rovelli and Vidotto \cite{rovellividotto} discussing the
topic. Closely related to these are a set of theories such that their
Feynman diagrammatics coincide with the diagrams that spin foams
produce \cite{groupfieldtheory}. This is analogous to what matrix
models do in $1+1$ dimensions. These field theories are called group
field theories and can be viewed as generalizations of matrix models
to four dimensions.

Summarizing, loop quantum gravity is now close to 30 years old. There
is a proposal for a theory of quantum gravity based on it that is
finite, anomaly free and well defined. It has proved challenging to
confirm if it contains the correct physics in the semiclassical
limit. However, steady progress in symmetry reduced models is
producing attractive results that may lead to connections with
experiment in the cosmological context and may open new lines of
attack on the problem of back reaction in black hole evaporation in
the spherically symmetric context. Important progress in these areas
is likely to happen in a few short years. It can fairly be said that
this quantum legacy of the ADM papers is a testimony of the profound
influence they keep on having in gravitational physics.

\section{Acknowledgements}
I wish to thank Ivan Agull\'o, Stanley Deser, Rodolfo Gambini, Charles
Misner and Javier Olmedo for comments. 
This work was supported in part by grant NSF-PHY-1305000, CCT-LSU, and
the Horace Hearne Jr. Institute for Theoretical Physics at LSU. 

\end{document}